\documentclass[aps,prb,twocolumn,floatfix,superscriptaddress]{revtex4}

\usepackage{amsmath,bm,latexsym,dcolumn,multirow}
\usepackage{graphicx} \graphicspath{{./figures/}}

\newcommand{\half}{\frac{1}{2}}
\newcommand{\grid}[1]{$#1{\times}#1{\times}#1$} 
\newcommand{\mcds}[2]{\multicolumn{1}{d}{#1\textsuperscript{#2}}}
\newcommand{\lamMP}{\multicolumn{1}{c}{$\lambda^{\rm (MP)}$}}
\newcommand{\lamLW}{\multicolumn{1}{c}{$\lambda^\Gamma$}}
\newcommand{\lamEX}{\multicolumn{3}{c}{$\lambda^{\rm expt}$}}

\newcommand{\aitops}{\textsc{ai{\footnotesize 2}ps}}
\newcommand{\feff}{\textsc{feff{\footnotesize 9}}}
\newcommand{\abinit}{\textsc{abinit}}

\begin{document}

\title{Cumulant expansion for phonon contributions to the electron 
        spectral function}

\author{S. M. Story}    \affiliation{Department of Physics, University of Washington 
                                     Seattle, WA 98195}

\author{J. J. Kas}      \affiliation{Department of Physics, University of Washington 
                                     Seattle, WA 98195}

\author{F. D. Vila}     \affiliation{Department of Physics, University of Washington 
                                     Seattle, WA 98195}

\author{M. J. Verstraete} \affiliation{Institut de Physique, Universit\'e de Li\`ege
                                       B-4000 Sart Tilman, Belgium}

\author{J. J. Rehr}     \affiliation{Department of Physics, University of Washington
                                     Seattle, WA 98195}


\date{\today}


\begin{abstract}
We describe an approach for calculations of phonon contributions to the electron 
spectral function, including both quasiparticle properties and satellites. The 
method is based on a cumulant expansion for the retarded one-electron Green's 
function and a many-pole model for the electron self-energy. The electron-phonon 
couplings are calculated from the Eliashberg functions, and the phonon density of 
states is obtained from a Lanczos representation of the phonon Green's function. Our
calculations incorporate {\it ab initio} dynamical matrices and electron-phonon 
couplings from the density functional theory code \abinit{}. Illustrative results are
presented for several elemental metals and for Einstein and Debye models with a range
of coupling constants. These are compared with experiment and other theoretical 
models. Estimates of corrections to Migdal's theorem are obtained by comparing with 
leading order contributions to the self-energy, and are found to be significant only
for large electron-phonon couplings at low temperatures.
\end{abstract}

\maketitle


\section{INTRODUCTION}

To first approximation, electronic and vibrational properties can be treated
separately in condensed matter due to the large mass ratio between electrons and 
ions, e.g., within the Born-Oppenheimer approximation. However, corrections to this
approximation, which depend on the strength of the electron-phonon interaction, are
of considerable importance both theoretically and experimentally.
Here we investigate the effects of electron-phonon interactions on the quasiparticle 
properties of electrons.  Due to such interactions, the electron energy levels 
$\varepsilon_k$ are not sharply defined, but have finite lifetimes characterized by
the electron self-energy $\Sigma$, which lead to broadening of the associated 
spectral function. Such effects are visible experimentally, e.g., in high resolution
ARPES spectra at low temperatures.\cite{cuketal}

In general, the electron spectral function is dominated by a sharp quasiparticle 
peak, but it can also exhibit satellites due to phonon excitations. According to 
Migdal's theorem,\cite{migdalJETP58} only the leading order electron-phonon 
interaction contributions to the self-energy are important, due to the large mass
ratio between electrons and nuclei. In that case, the electron self-energy can be 
approximated by the simplest diagram, and vertex corrections can be neglected. This
approximation has been investigated in detail 
\cite{engelsbergPR63,bernardiNL13,qiuPRL13,hybertsenPRB86,onidaRMP02}
and extended to finite temperature, e.g., by Allen. \cite{allenSSP82} 
The Migdal approximation is analogous to the $GW$ approximation of Hedin 
\cite{hedinPSC99} for electrons coupled to plasmons where $G$ is the electron Green's
function and $W$ the screened Coulomb interaction. Since a similar formalism applies 
to phonons, electron-hole pairs and other neutral bosonic excitations, we will refer
to this leading order diagram as the GW approximation. The GW theory leads to
spectral functions with a quasiparticle peak and two satellite features originating
from single-boson excitations, one on each side of the main quasiparticle peak. 

In contrast to the GW approximation, however, systems of electrons coupled to neutral
bosonic excitations generally exhibit multiple satellites, as observed in 
photoemission experiments. \cite{steinerTAP79,offiPRB07} Moreover, the GW
approximation is known to be unsatisfactory for describing satellite structures, as
the satellite peaks typically appear at the wrong energies and with the wrong 
intensities compared to experiment. Thus, it is of interest to investigate possible 
corrections to Migdal's theorem, i.e., the effects of higher order terms in an 
expansion in powers of the electron-phonon coupling.\cite{gunnarssonPRB94} One 
approach to this end is to investigate contributions to the self-energy from the
vertex function $\Gamma$, as in the formal identity $\Sigma = iGW\Gamma$. However, 
direct calculations of $\Gamma$ have been formidably challenging, and there has been
little progress along these lines.  An attractive alternative that overcomes some of
the shortcomings of GW is provided by the cumulant expansion,
\cite{kuboJPSJ62,gunnarssonPRB94,kas2014} 
which is an exponential representation of the electron Green's function in the time 
domain. The cumulant expansion is exact for the case of a deep core-level coupled to
bosons, and generalizations have been developed for valence electrons coupled to 
plasmons. \cite{hedinPSC80,aryasetiawanPRL96} 
The approach has been applied with considerable success in many cases, ranging from
multiple plasmon satellites in photoemission\cite{guzzoPRL11} to dynamical mean field
theory.\cite{casulaPRB12} Nevertheless, the conventional approach based on the 
time-ordered Green's function is only strictly applicable for the hole- \emph{or} 
particle-branch of the spectral function depending on whether the state is above or 
below the Fermi level. This limitation is particularly problematic in systems with
particle-hole symmetry, such as electrons coupled to phonons. To overcome this 
difficulty, we utilize here the recently developed retarded cumulant (RC) approach,
which is based on a particle/hole cumulant and a \emph{retarded} Green's function
formalism.\cite{kas2014} A further goal of the present work is to develop a practical
approach for calculations of phonon contributions to properties of condensed matter.

The remainder of this paper is organized as follows. In Sec.~\ref{sec:theory}, we
describe the retarded cumulant expansion method and many-pole model self-energy used
to calculate phonon contributions to the electron spectral function. 
Sec.~\ref{sec:implement} gives details on how this method is implemented 
computationally with our workflow tool \aitops{}. Finally, our resents are presented
in Sec.~\ref{sec:results}, and Sec.~\ref{sec:summary} contains a summary and 
conclusions.


\section{THEORY AND METHODOLOGY} \label{sec:theory}

In this section, we briefly summarize the GW and RC approximations for calculations 
of the electron spectral function in systems linearly coupled to phonons. As usual,
the Hamiltonian for the electron-phonon system is represented as
\begin{equation}
H = \sum_k \varepsilon^0_k c_k^{\dagger} c_k + \sum_q \omega_q a_q^{\dagger} a_q
+ \sum_{kk'q} V^q_{kk'} (a_q + a_q^{\dagger}) c_k^{\dagger} c_{k'},
\end{equation}
where $k$ denotes the electron levels and $q$ the phonon modes with bare energies
$\varepsilon^0_k$ and $\omega_q$ respectively, $V^q_{kk'}$ are the electron-phonon
matrix elements, and $c_k$ ($c_k^{\dagger}$) and $a_q$ ($a_q^{\dagger}$) are the 
electron and phonon destruction (creation) operators. In this paper, we use atomic
units $e=\hbar=m=1$ and $k_B$ = 0.086173 meV/K. At low temperatures, the electrons
are nearly degenerate with Fermi energy $\varepsilon_F$ and $\omega_q\ll\varepsilon_F
\ll\omega_p$, where $\omega_p$ is the dominant plasmon excitation energy, which is
typically several eV. Thus, for our purposes here, the density of electron states 
near $\varepsilon_F$ replaced by a constant, which we assume is non-vanishing. The
generalization to insulators or molecular systems with discrete spectra near
$\varepsilon_F$ is straightforward, but will not be treated here.

\subsection{GW spectral function}

Schematically, the GW approximation for the self-energy is given by $\Sigma=iGW$,
where $G$ is the one-electron Green's function and $W$ is an approximation for the 
screened Coulomb interaction. Within GW, the usual strategy is to calculate the
spectral function $A_k(\omega)$ from the imaginary part of the one-electron Green's
function in frequency space,\cite{engelsbergPR63}
\begin{equation}
\begin{aligned}
 \label{eq:gw}
  G_k\left(\omega\right)&=\frac{1}{\omega-\varepsilon^0_k-\Sigma_k
    \left(\omega\right)} \\
  A_k\left(\omega\right) &=\frac{1}{\pi}\left|{\rm Im}\,G_k(\omega)\right| \\
    &= \frac{1}{\pi}\frac {|{\rm Im}\, \Sigma_k(\omega)|}
    {|\omega-\varepsilon_k^0- {\rm Re}\, \Sigma_k(\omega)|^2
    + |{\rm Im}\, \Sigma_k(\omega)|^2 }.
\end{aligned}
\end{equation}
The spectral function is comprised of two main features---a dominant quasiparticle 
peak at $\omega=\varepsilon_k=\varepsilon_k^0+\Sigma_k$ with width ${\rm Im}\,
\Sigma_k$ and phonon satellites at $\omega=\varepsilon_F\pm\omega_q$, consistent with 
Ref.~\onlinecite{engelsbergPR63}. Other physical properties such as the quasiparticle
lifetime and energy levels can be obtained from the properties of $A_k(\omega)$ and
$\Sigma_k(\omega)$.

\subsection{RC spectral function}

As noted in the introduction, the conventional time-ordered cumulant expansion must
be generalized to treat cases with particle-hole symmetry, such as phonon excitations
in metals. \cite{gunnarssonPRB94} Our treatment is based on the RC formalism which is
discussed in detail by Kas et al.\cite{kas2014} For a degenerate Fermi system in the
absence of plasmons, the RC representation of the retarded one-particle Green's 
function is
\begin{equation}
\begin{aligned}
\label{eq:greensRC}
  G^R_k\left(t\right)&= G_k^{0,R}(t) e^{C^R_k(t)} \\
  G_k^{0,R}\left(t\right)&= -i\,e^{-i\varepsilon^0_kt}\theta\left(t\right),
\end{aligned}
\end{equation}
where $C^R_k(t)$ is the cumulant as described below. Formally, the spectral function
is obtained from a Fourier transform
\begin{equation}
  A_k\left(\omega\right)={\rm Im}\,\int_{-\infty}^\infty\frac{dt}{2\pi}\; i 
  e^{i\omega t}G_k^R\left(t\right).
\end{equation}
The retarded particle/hole cumulant $C^R_k(t)$ is then approximated by the second 
order (in electron-phonon coupling) cumulant diagram \cite{kas2014}
\begin{equation}
\begin{aligned}
\label{c2k}
  C^R_k(t) &\approx C^R_{2,k}\left(t\right) \\
   &= i e^{i\varepsilon^0_k t}
  \int_{-\infty}^\infty\frac{d\omega}{2 \pi}e^{-i\omega t} 
  \left[G^{0,R}_k\left(\omega\right)\right]^2\Sigma^R_k\left(\omega\right).
\end{aligned}
\end{equation}
This diagram is conveniently evaluated in frequency space \cite{kas2014} and can be
expressed in terms of the imaginary part of the $G^0W^0$ boson excitation spectrum 
$\beta_k(\omega)$ as
\begin{equation}
 \label{eq:Cumulant}
  C_k^R(t)=\int_{-\infty}^\infty d\omega\,\beta_k(\omega)\,
         \frac{e^{i\omega t}-i\omega t-1}{\omega^2}, \\
\end{equation}
where $\beta_k(\omega)$ is obtained from the GW self-energy
\begin{equation}
\label{eq:beta}
  \beta_k(\omega) = \frac{1}{\pi}\left|{\rm Im}\,
    \Sigma_k\left(\omega+\varepsilon^0_k\right)\right|.
\end{equation}
Consequently the ingredients in the RC are similar to those in GW and and hence the
RC is no more difficult to calculate than the GW approximation. In contrast to the 
conventional time-ordered cumulant expansion, which only contains frequencies within
the particle- or hole branches, the retarded cumulant in Eq.~(\ref{eq:Cumulant}) 
contains \emph{all} frequencies, and explicitly builds in the particle-hole symmetry 
desired for phonons. Also, due to the behavior of the essentially dispersionless
self-energy $\Sigma_k(\omega)$, (Fig.\ \ref{fig:SE}), multiple phonon satellites may
exist with the cumulant expansion, as peaks at integral multiples of $\omega_E$ on
both sides of the Fermi energy $\epsilon_F$. This is in contrast to the case with
plasmons, where the satellites appear at multiples of $\omega_p$ from the
quasiparticle peak at $\epsilon_k$.

\subsection{Many-pole GW self-energy}

\begin{figure}[t]
 \includegraphics[width=\columnwidth]{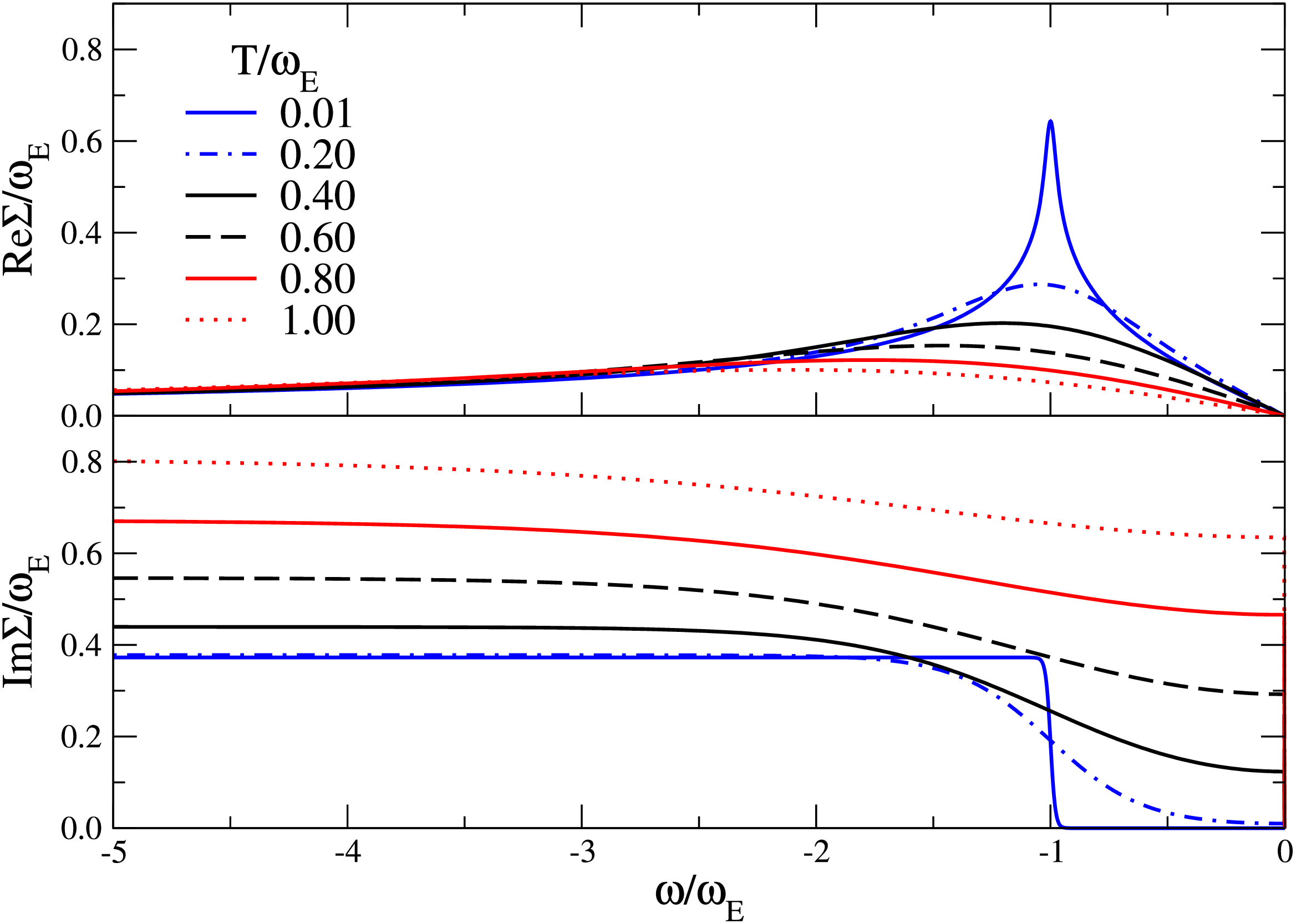}
 \caption{(color online) Real (top) and imaginary (bottom) parts of the self-energy
   $\Sigma_k(\omega)$ in Eq.~(\ref{eq:SE}) using the Einstein model for Cu, where 
   $\omega_E$ = 21.6\ meV = 251 K (see text). Positive $\omega$ is not shown, as 
   ${\rm Re}\,\Sigma$ and ${\rm Im}\,\Sigma$ and can be obtained from the parity of
   $\Sigma_k(\omega)$ versus $\omega$.}
 \label{fig:SE}
\end{figure}

The dominant ingredient in the RC is the $G^0W^0$ boson excitation spectrum
$\beta_k(\omega)$, which is general for any given self-energy, but we will focus on
a self-energy model appropriate for phonons. Here we have adapted the 
finite-temperature Einstein model for phonons,
\cite{grimvall81,allenSSP82,eigurenPRL08} where the self-energy is represented as a
sum over Einstein modes. For a single mode with Einstein frequency $\omega'$, the GW
self-energy at finite temperature $T$ (with unit coupling) is given by
\cite{allenSSP82,eigurenPRL08}
\begin{equation}
\begin{split}
 \label{eq:einSE}
 &  \Sigma^{\rm E}\left(\omega,\omega',T\right) 
  =  -i\pi\, \left[n\left(\omega'\right)+\half\right]  + \\
 & + \half\Psi\left(\half+i\frac{\omega'-\omega}{2\pi T}\right) 
   -\half\Psi\left(\half-i\frac{\omega'+\omega}{2\pi T}\right) ,
\end{split}
\end{equation}
where $n(\omega)$ is the Bose-Einstein distribution and $\Psi(z)$ is the digamma
function. The electron-phonon coupling constants in the model are represented in 
terms of the Eliashberg function $\alpha^2 F_k(\omega)$.  The self-energy to be used
for $\beta_k(\omega)$ in Eq.~(\ref{eq:beta}) is then \cite{grimvall81,allenSSP82}
\begin{equation}
  \label{eq:SE}
  \Sigma_k\left(\omega,T\right)=\int\,d\omega'\,2\Sigma^{\rm E}\left(\omega,
  \omega',T\right)\alpha^2F_k\left(\omega'\right).
\end{equation}
We emphasize that the form of the self-energy in Eq.~(\ref{eq:SE}) is strictly 
appropriate only for cases where the band width of electron states near the Fermi
energy is large compared to characteristic phonon energies $\omega$, and will not 
work for sharp band features. This is the case for valence states in metals and in 
many semi-metals, semiconductors, and insulators, but becomes questionable in the 
case of small molecules and core level states. Thus in the present work, we focus 
only on a selection of metallic systems with a range of electron-phonon couplings.
As an example, Fig.~\ref{fig:SE} shows the real and imaginary parts of the 
self-energy calculated using Eq.~(\ref{eq:SE}) from coupling to a single Einstein 
mode, i.e., an Einstein model for the phonon spectrum in Cu.

\begin{figure}[t]
  \includegraphics[width=\columnwidth]{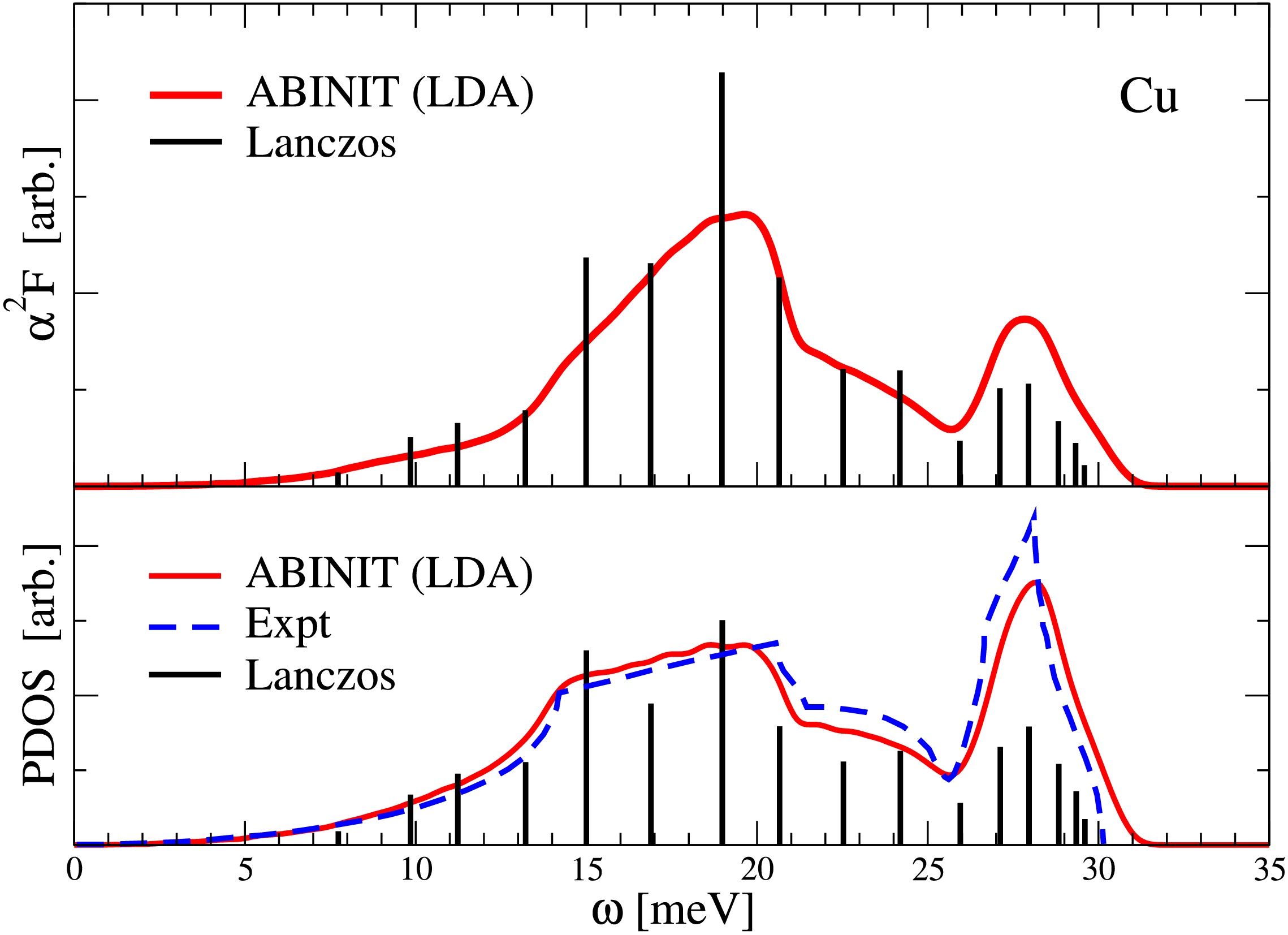}
  \caption{(color online) Eliashberg function (top) $\alpha^2F(\omega)$ and total
    density of modes (bottom) for Cu at the Fermi level $k=k_F$ obtained from 
    \abinit{} with our many-pole approximations $\alpha^2_i,F_i$ calculated by 
    the Lanczos inversion tools in \feff{} (see text). Experimental PDOS taken from
    Ref.~\onlinecite{nicklowPR67}. Calculated frequencies have been scaled to match
    the peak frequency with experiment.}
   \label{fig:a2f}
\end{figure}

Moreover, for computational simplicity, it is convenient to use a many-pole model for
the self-energy,\cite{engelsbergPR63} analogous to the plasmon-pole model of Hedin 
and Lundqvist.\cite{hedinSSP70,lundqvistPKM67single,lundqvistPKM67numerical,kasPRB07}
The integration over the phonon frequencies $\omega'$ in Eq.~(\ref{eq:SE}) can be
replaced by a discrete sum over a sufficiently large number of poles without 
significant loss of accuracy. For the electron-phonon couplings 
$\alpha^2F_k(\omega)$, we employ a pole model similar to that used for the dielectric
function in Ref.~\onlinecite{kasPRB07}. Generally, $\alpha^2F_k(\omega)$ depends on 
both $k$ and $k'$ through the electron-phonon matrix elements $g^q_{kk'}$.
\cite{mcmillanPR68,grimvallPKM70,allenPRB72,savrasovPRB96}
However, since the phonon spectra involve energies very close to $\varepsilon_F$, it 
is sufficient for our purposes here to use the Eliashberg function averaged over the
Fermi surface
\begin{equation}
\begin{aligned}
  \alpha^2F(\omega)&=\frac{1}{2\pi N(\varepsilon_F)}\sum_q\sum_{k,k'\approx
    \varepsilon_F}\left|g^q_{kk'}\right|^2\delta(\omega-\omega_q) \\
  g^q_{kk'}&=\sum_{\sigma\sigma'}\left\langle\psi_{\sigma',k'}\right|\delta
    V^q_{kk'}\left|\psi_{\sigma,k}\right\rangle,
\end{aligned}
\end{equation}
where $k=k'+q\approx\varepsilon_F$, $N(\varepsilon_F)$ is the bare density of states
at the Fermi level, and $\sigma$ denotes spin states. 
Typically, the $\alpha^2F(\omega)$ spectrum is rather similar to the total phonon 
density of states (PDOS) $F(\omega)$ in the system (see Fig.~\ref{fig:a2f}), for 
which an efficient many-pole Lanczos representation has been 
developed,\cite{vilaPRB07}
\begin{equation}
 F^{\rm MP}(\omega) =  \sum_i F_i \delta(\omega-\omega_i).
\end{equation}
Thus, a many-pole representation of $\alpha^{2}F^{\rm MP}$ can be constructed 
similarly,
\begin{equation}
  \alpha^2F^{\rm MP}\left(\omega\right)=\sum_i\alpha^2_iF_i\delta\left(\omega-
  \omega_i\right).
 \label{eq:eliMP}
\end{equation} 
Here the amplitudes
\begin{equation}
  \alpha^2_i=\alpha^2 F(\omega_i)/F(\omega_i)
  \label{eq:couplings}
\end{equation} 
represent the discretized electron-phonon couplings. A 16-pole representation of the
copper Eliashberg function is shown in Fig.~\ref{fig:a2f}. Finally, an effective or
mean electron-phonon coupling constant $\lambda$ can be defined, which is related to
the first inverse frequency moment of the Eliashberg function \cite{allenAP99}
\begin{equation}
\label{eq:lambda}
  \lambda=2\int_0^\infty \dfrac{d\omega}{\omega}\alpha^2F\left(\omega\right)
         \approx 2\sum_i\dfrac{\alpha^2_iF_i}{\omega_i}.
\end{equation}
This quantity provides a dimensionless characterization of the strength of 
electron-phonon coupling in a given material.


\section{IMPLEMENTATION} \label{sec:implement}

The calculations of phonon properties presented here were carried out using \aitops{}
({\it ab initio} DFT to Phonon Spectra),\cite{ai2ps} a workflow tool we have 
developed that links density functional theory electronic structure codes, \abinit{} 
in this case, \cite{gonzeCMS02,gonzeZK05} 
to the vibrational properties module of real-space Green's function code \feff{}.
\cite{feff9} \aitops{} can be used to calculate phonon properties such as 
Debye-Waller factors in x-ray spectra. The modular interface automatically 
coordinates the desired workflow.
Briefly, for our purposes here, \aitops{} uses \abinit{} to generate a set of 
real-space symmetry-inequivalent blocks of the lattice dynamical matrix (DM),
which are used to calculate the many-pole PDOS $F^{\rm MP}(\omega)$.\cite{vilaPRB07} 
The code \abinit{} also yields both $F(\omega)$ and $\alpha^2F(\omega)$, which are
used to calculate the couplings $\alpha^2(\omega)$ using Eq.~(\ref{eq:couplings}).
Since Eq.~(\ref{eq:SE}) is restricted to energies near the Fermi level, this 
presently excludes any $k$-dependent features in the spectral functions presented in
the current study.
The \abinit{} calculations used Troullier-Martins/Fritz Haber Institut LDA
pseudopotentials, and an energy cutoff of 50 Hartrees; for convergence of  
$\alpha^2F(\omega)$, a \grid{32} Monkhorst-Pack $k$-point grid was found to be 
necessary.  For the metallic systems discussed here, the occupation numbers were
smeared with the Methfessel and Paxton scheme\cite{methfesselPRB89} with a broadening
parameter of 0.025. 
Runtimes were dominated by the \abinit{} portion of the workflow. Using 160 AMD 
Opteron 6128 (800 MHz) cores spread across ten nodes, the runtime for one set of 
parameters is split roughly 99\% ($\sim$200 minutes) \abinit{} for the coupling 
constants and 1\% ($\sim$2 minutes) \feff{} for the vibrational properties.
Calculations of the spectral function $A_k(\omega)$ were parameterized by the
quasiparticle energy $\varepsilon_k=\varepsilon_k^0+{\rm Re}\Sigma_k(
\varepsilon_k^0)$ instead of $\varepsilon_k^0$ (see Eq.~(\ref{eq:greensRC}) and 
(\ref{eq:beta})). This further simplified the calculation by removing self-energy
shifts.


\section{Results and Discussion} \label{sec:results}

In this section, we present illustrative results for several elemental metals and for
Einstein and Debye models with a range of electron-phonon couplings over a range of
temperatures and energies for both the RC and GW methods.

\subsection{Einstein model}

\begin{figure}[t]
  \includegraphics[width=\columnwidth]{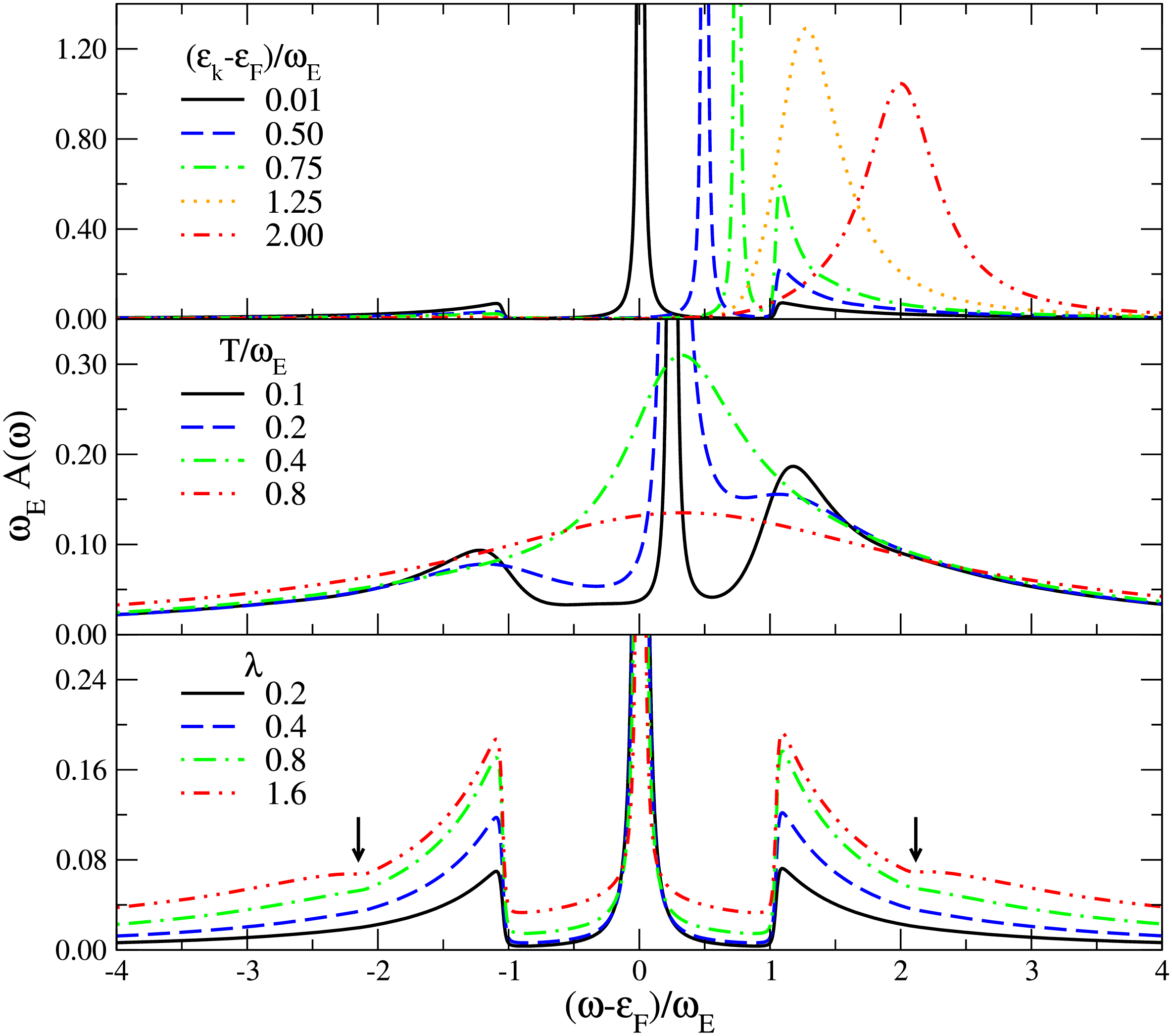} 
  \caption{(color online) Spectral function for the Einstein model using the RC 
    method, where $\omega_E$ is the Einstein energy. Top: varying quasiparticle 
    energy for low temperature and weak coupling ($T=0.01\ \omega_E$, $\lambda=0.2$),
    middle: varying temperature near the Fermi energy and with medium coupling 
    ($\varepsilon_k-\varepsilon_F=0.25\ \omega_E$, $\lambda=1.0$), bottom: varying
    electron-phonon coupling constant at low temperature near the Fermi energy 
    ($\varepsilon_k-\varepsilon_F=T=0.01\ \omega_E$).}
  \label{fig:ein}
\end{figure}

As a first example, we consider the Einstein model self-energy $\Sigma^{\rm E}$, 
i.e., using the single-pole (zeroth-order Lanczos) approximation for the Eliashberg
function,
\begin{equation}
  \alpha^2F\left(\omega\right)=\alpha^2\left(\omega_E\right)\delta\left(\omega-
  \omega_E\right),
\end{equation} 
where $\omega_E$ is the Einstein frequency. For realistic systems, the value of 
$\omega_E$ is taken to be the centroid of the PDOS provided by the \abinit{} 
calculation. As an example, we present results for an Einstein model with $\omega_E=
21.6$ meV (251 K) representative of Cu metal in Fig.~\ref{fig:ein}. Note that phonon
satellites in the spectral function are visible only for quasiparticle energies small
compared to phonon frequencies $\varepsilon_k<\omega_E$, and very low temperatures 
($\sim10$ K), as seen in the top two panels of Fig.~\ref{fig:ein}. 
For the Einstein model, the mean coupling constant $\lambda$ in Eq.~(\ref{eq:lambda})
is simply $2\alpha^2/\omega_E$, so we can artificially ramp up the coupling by
manually setting the value of $\alpha^2$. Typically, metals have coupling constants
$\lambda$ that range from roughly 0.1 to 1.7,\cite{savrasovPRB96} so we will focus on
that range. The satellites become larger as $\lambda$ is increased (third panel), and
for $\lambda\approx 1.6$, a weak second phonon satellite becomes apparent at $\omega=
\varepsilon_F\pm\ 2\omega_E$. The relative weakness of the 2nd satellite even at 
$\lambda=1.6$ suggests Migdal's theorem is valid to high accuracy for typical metals,
apart from corrections close to the Fermi energy at very low temperatures.

\subsection{Debye model}

\begin{figure}[t]
  \includegraphics[width=\columnwidth]{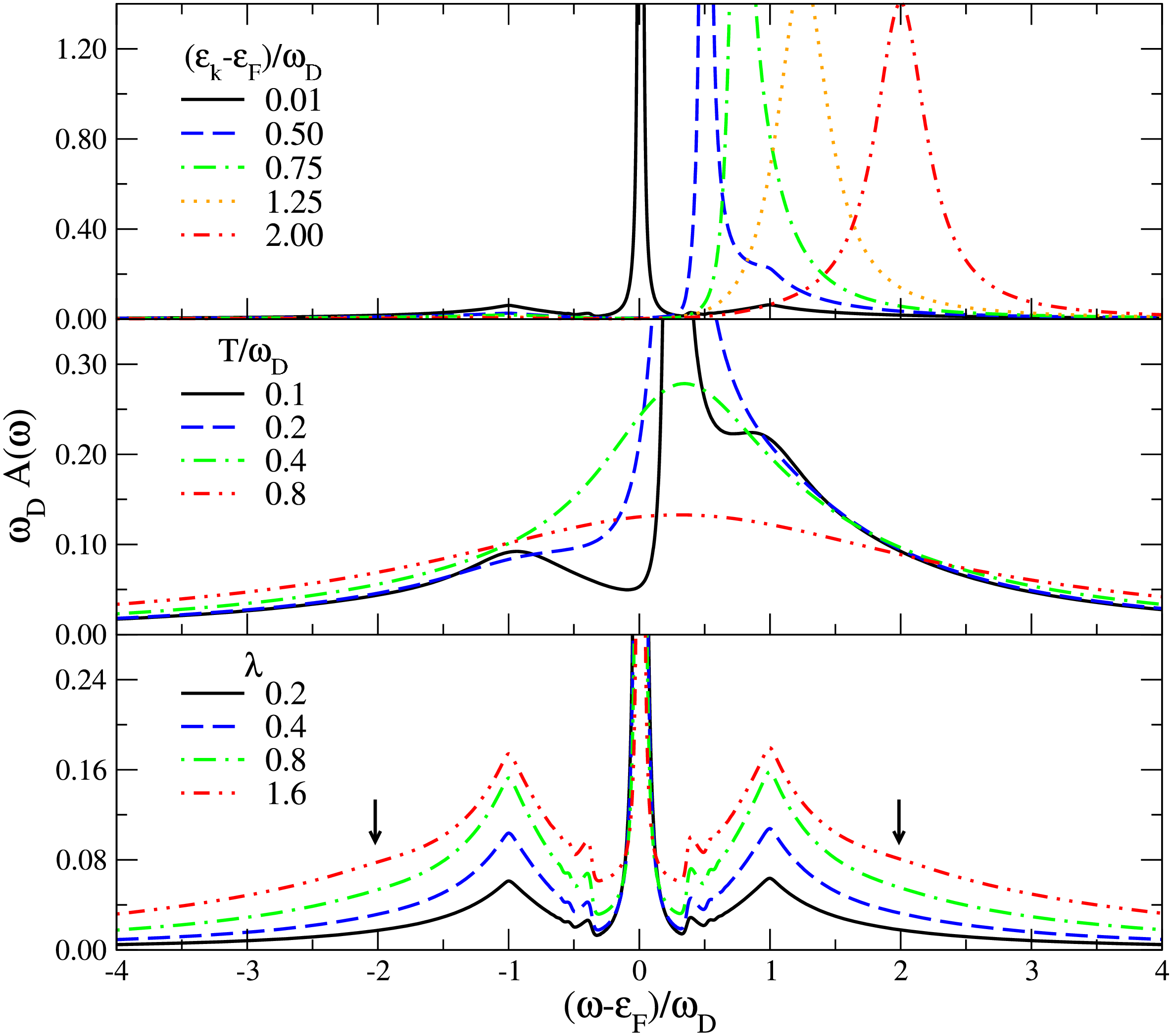} 
  \caption{(color online) Spectral function for the Debye model using the RC method,
    where $\omega_D$ is energy corresponding to the Debye temperature. Top: varying
    quasiparticle energy for low temperature and weak coupling ($T=0.01\ \omega_D$,
    $\lambda=0.2$), middle: varying temperature near the Fermi energy and with medium
    coupling ($\varepsilon_k-\varepsilon_F=0.25\ \omega_D$, $\lambda=1.0$), bottom: 
    varying electron-phonon coupling constant at low temperature near the Fermi 
    energy ($\varepsilon_k-\varepsilon_F=T=0.01\ \omega_D$).}
  \label{fig:deb}
\end{figure}

For comparison, we show similar results using the Debye model PDOS converted to
a many-pole form in Fig.~\ref{fig:deb}, with quantities expressed in terms of the
Debye temperature for copper $\Theta_D=315\textrm{ K}=27.1\ \textrm{ meV}=\omega_D$.
Overall, the Debye model shows trends quite similar to the Einstein model. However,
the phonon satellites are not as sharply peaked, and the satellites at $2\omega_D$
are barely visible at the same scale for large couplings $\lambda\sim 1.6$. Note 
that artifacts of the many-pole model can be seen in the spectral functions as small
peaks near the Fermi energy (third panel), though these are negligible compared to 
the phonon satellites.

\subsection{Comparison of RC and GW}

\begin{figure}[h]
  \includegraphics[width=\columnwidth]{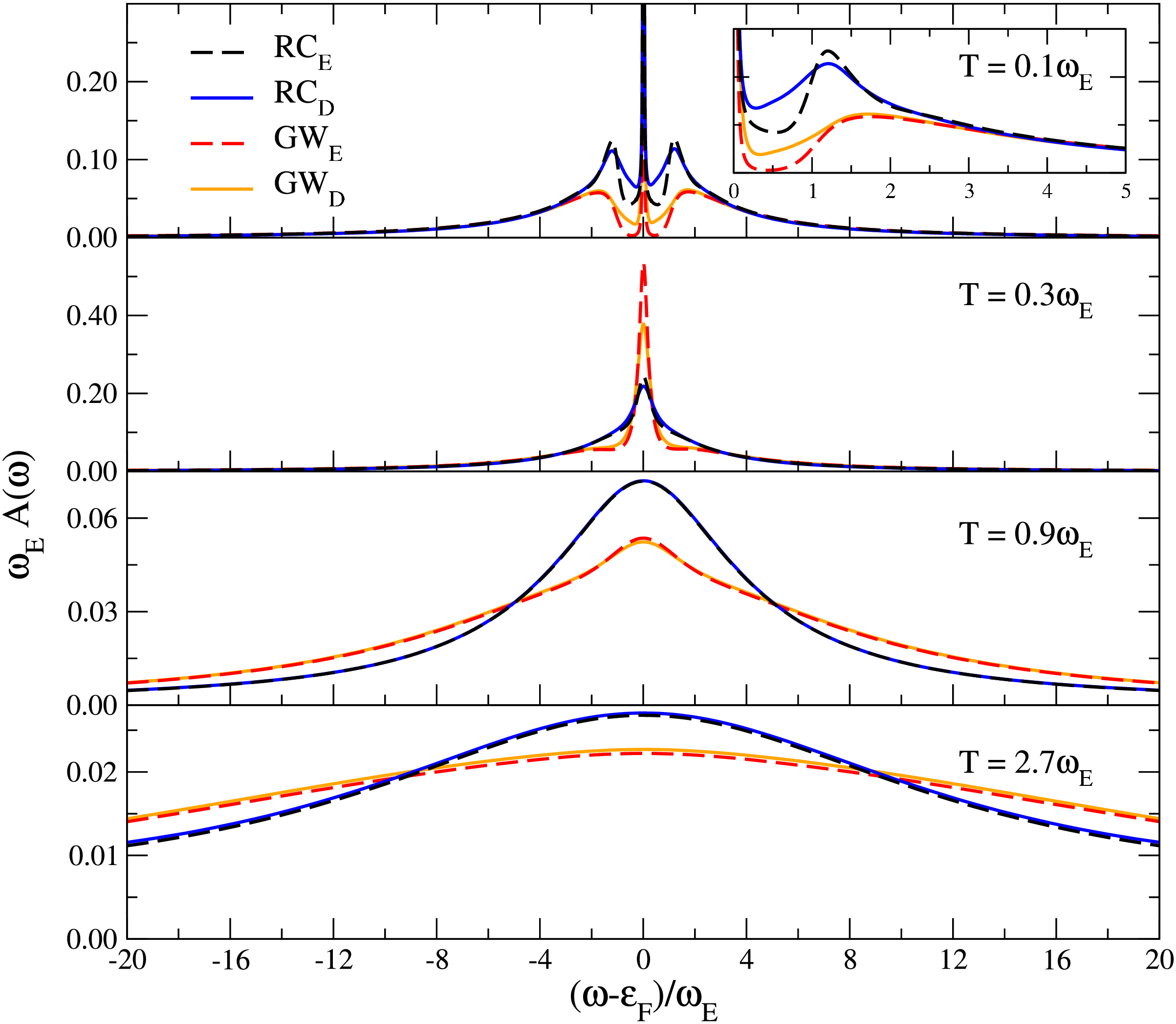} 
  \caption{(color online) Comparison of spectral function from the RC and GW methods
    using the Einstein and Debye models at strong coupling (i.e., $\lambda=1.6$) near
    the Fermi level ($\varepsilon_k= \varepsilon_F+0.01\ \omega_E$) for various 
    temperatures, where $\omega_D\approx1.3\ \omega_E$. Inset: enlarged view of 
    phonon satellites seen in the top panel. Note the satellite centroid for the GW
    method is further out, and the kink in the RC satellite at 2$\omega_E$.} 
  \label{fig:gw} 
\end{figure}

We note that the electron spectral function near the Fermi level $k=k_F$ is generally
nearly symmetrical due to particle-hole symmetry, and is sensitive to phonon 
correlations beyond GW at strong electron-phonon coupling, as illustrated by the 
significant deviation of RC from GW seen in Fig.~\ref{fig:gw}. Thus, it is useful to 
compare the RC and GW methods in this limit, especially since the differences 
characterize corrections to the GW approximation due to vertex effects. 
Fig.~\ref{fig:gw} shows that the two methods differ significantly at strong couplings 
and low temperatures compared to the Debye or Einstein temperature (see Table 
\ref{tab:weights} for distribution of spectral weight). The RC method gives larger 
satellite weights, with a strong first satellite peak at $\omega_E$ and a slight kink
at $2\omega_E$ (see Inset to Fig.~\ref{fig:gw}). However, the differences between the
two methods diminish as the temperature is increased towards room temperature. 

\begin{table}[h]
\caption{Comparison of weights for the quasiparticle peak, hole satellite, and particle 
    satellite ($Z_k,w_h,w_p$ respectively) using the RC and GW methods, done for the 
    Einstein/Debye models at large coupling ($\lambda$ =1.6) and several elemental 
    metals near the Fermi level at low temperature ($\varepsilon_k=\varepsilon_F+
    0.01\ \omega_{E,\rm Cu}$ = 0.216 meV, $T=0.1\ \omega_{E,\rm Cu}$ = 25.1 K).}
\label{tab:weights}
\begin{ruledtabular}
\begin{tabular}{l l c c c c}
                    &          &  $Z_k$  &  $w_h$  &  $w_p$  & $\lambda$ \\
\multirow{7}{*}{RC} & Einstein &   0.19  &   0.39  &   0.42  &    1.60   \\
                    & Debye    &   0.18  &   0.39  &   0.43  &    1.60   \\
                    & V        &   0.29  &   0.34  &   0.37  &    1.17   \\
                    & Nb       &   0.31  &   0.33  &   0.36  &    1.08   \\
                    & Pb       &   0.35  &   0.31  &   0.34  &    0.95   \\
                    & Ta       &   0.37  &   0.30  &   0.33  &    0.91   \\
                    & Cu       &   0.85  &   0.07  &   0.08  &    0.16   \\
\\                                                                       
\multirow{7}{*}{GW} & Einstein &   0.38  &   0.31  &   0.31  &    1.60   \\
                    & Debye    &   0.37  &   0.31  &   0.32  &    1.60   \\
                    & V        &   0.45  &   0.27  &   0.28  &    1.17   \\
                    & Nb       &   0.46  &   0.27  &   0.27  &    1.08   \\
                    & Pb       &   0.49  &   0.25  &   0.26  &    0.95   \\
                    & Ta       &   0.50  &   0.24  &   0.26  &    0.91   \\
                    & Cu       &   0.86  &   0.07  &   0.07  &    0.16   \\
\end{tabular}
\end{ruledtabular}
\end{table}

\subsection{Selected metals: Cu, Nb, Pb, Ta, and V}

\begin{figure}[b]
  \includegraphics[width=\columnwidth]{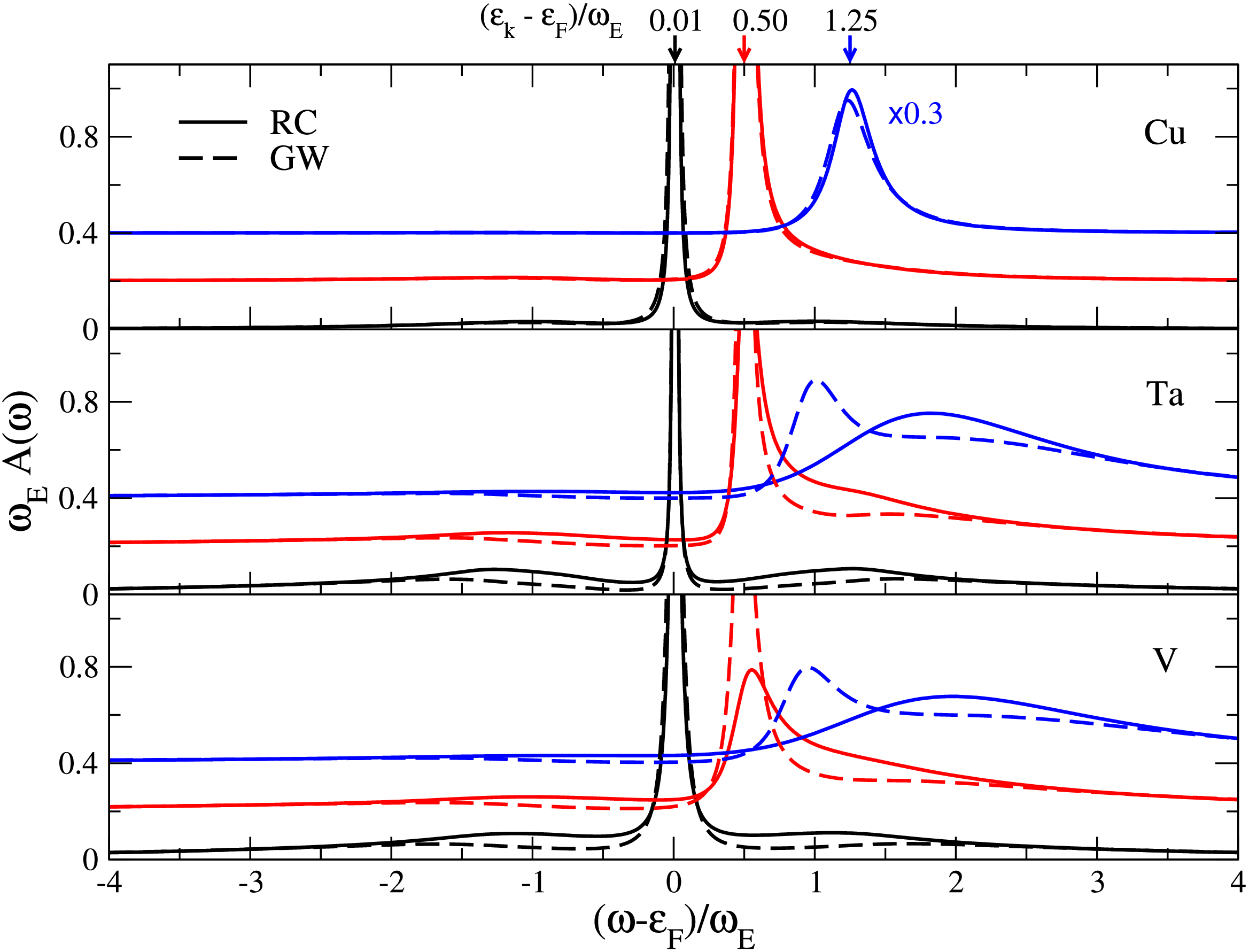}
  \caption{(color online) Comparing the spectral functions of the RC and GW methods
    for select metals at low temperature and three quasiparticle energies $\varepsilon_k$ for
    $(\varepsilon_k-\varepsilon_F)/\omega_E=0.01,0.5,1.25$ (bottom, middle, top
    vertically offset curves respectively in each panel, with arrows indicating 
    corresponding location along the horizontal axis). For Cu, Ta, and V, $\omega_E$
    = 21.6, 15.0, 24.1 meV respectively and $T=0.15\ \omega_E=$ 37.65, 17.4,
    42.0 K respectively.
    The spectral function for Cu with $\varepsilon_k-\varepsilon_F=1.25\ \omega_E$ 
    has been scaled vertically, as indicated.
    }
  \label{fig:metals}
\end{figure}

Next, we present results for the spectral function for a few elemental solids 
(Fig.~\ref{fig:metals}) representative of a range of electron-phonon couplings (See
Table \ref{tab:lambda}). To obtain the full spectral function for these materials,
we use the many-pole representation of Eq.~(\ref{eq:eliMP}) as calculated by \feff{},
as shown in Fig.~\ref{fig:a2f} for copper. 
The results for these metals follow similar trends with the RC and GW methods that we
saw with our results for the Einstein and Debye models. Copper, which has a 
relatively weak coupling ($\lambda \sim$ 0.1), displays near agreement between the RC
and GW methods. Tantalum and vanadium, on the other hand, have medium to strong 
couplings, respectively, and show significant differences between the two methods. 
Most noticeably, for increasing quasiparticle energies, both the distribution of 
weight between the quasiparticle and satellites and the location of these peaks 
disagree significantly, possibly enough to be noticeable experimentally. However,
these differences can only be seen at low temperatures ($\sim50$ K). Even with the 
strongest coupling, vanadium does not show multiple phonon satellites, indicating 
Migdal's theorem is valid to high accuracy for phonons in these materials.

\subsection{Comparison with experiment}

\begin{figure}[b]
  \includegraphics[width=\columnwidth]{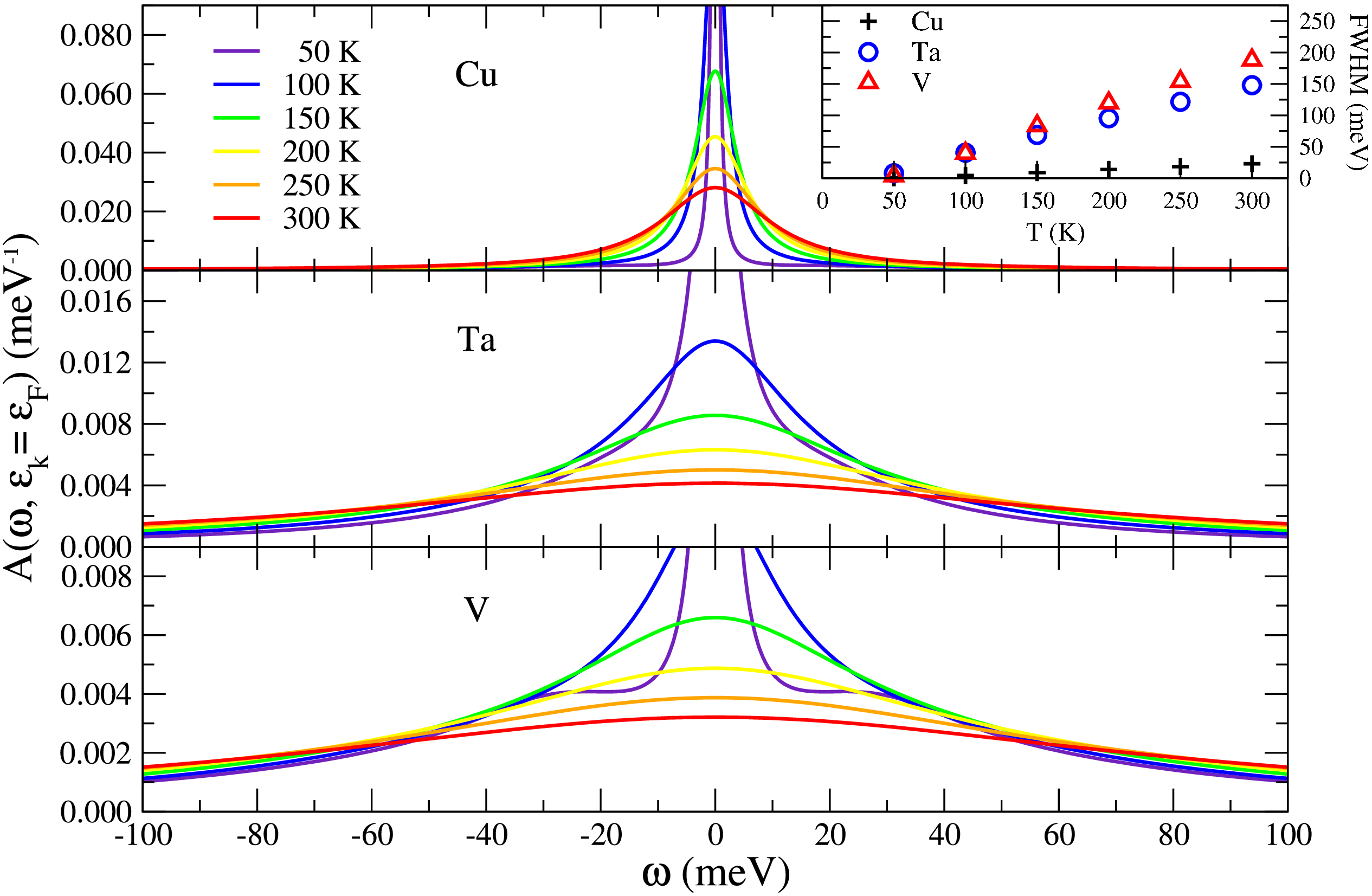}
  \includegraphics[width=\columnwidth]{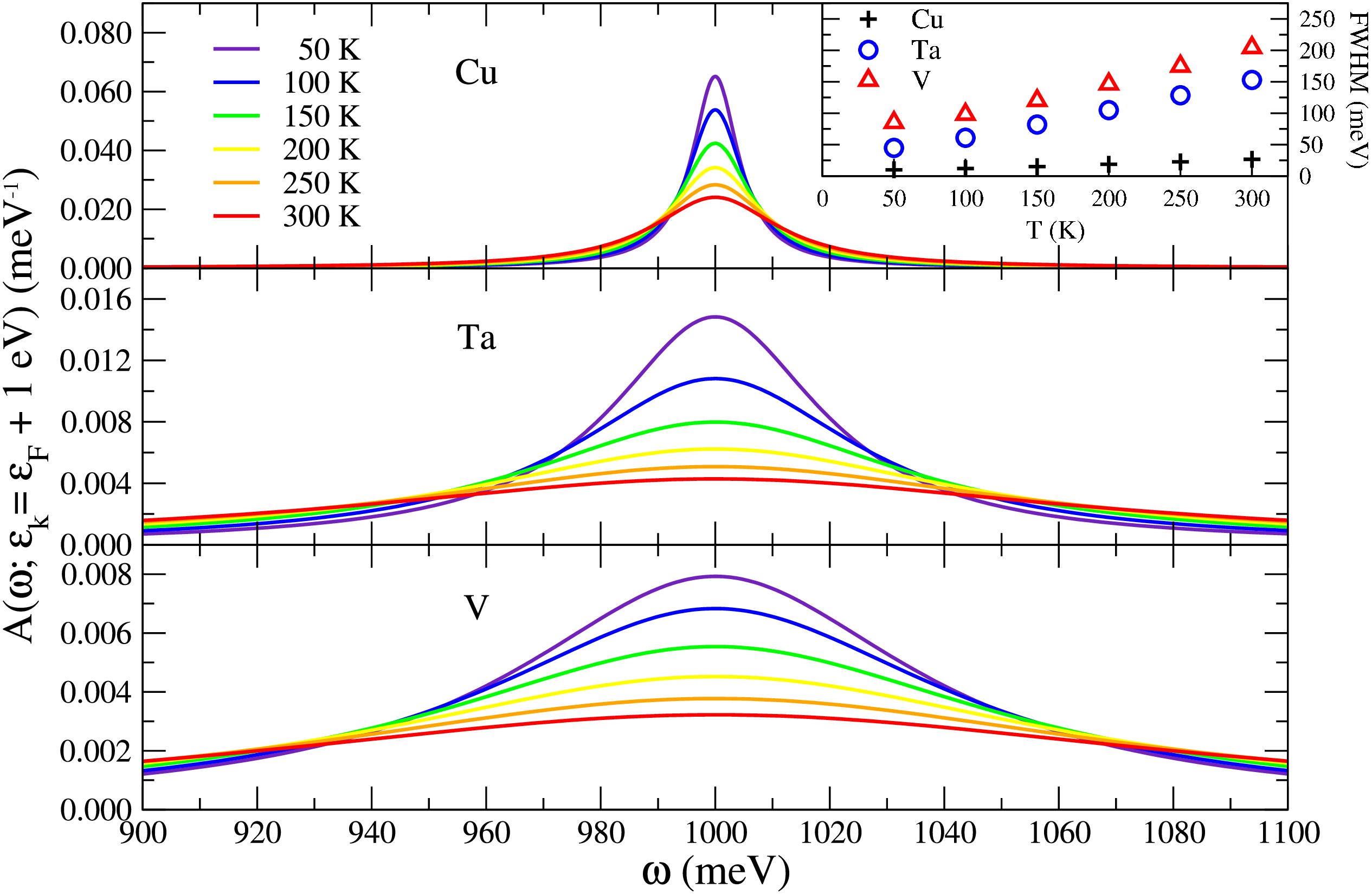}
  \caption{(color online) Behavior of the quasiparticle peak versus temperature at 
    the Fermi energy $\varepsilon_k=\varepsilon_F$ (top), and at moderate 
    quasiparticle energy $\varepsilon_k=\varepsilon_F+1.0$ eV (bottom) for Cu, Ta, 
    and V. Insets: Behavior of the quasiparticle widths $\Gamma$ (FWHM) versus 
    temperature, illustrating how the width of the spectral function grows linearly
    with $T$ according to $\Gamma\sim2\pi\lambda k_BT$. This relation can be used to
    estimate the electron-phonon coupling strength $\lambda$ (see Table 
    \ref{tab:lambda}). Note that the widths at the Fermi energy are significantly 
    reduced at low temperatures compared to those at $\varepsilon_k=\varepsilon_F$,
    due to the effect of phonon satellites on the distribution of spectral weight.}
  \label{fig:fwhm}
\end{figure}

\begin{table}[t]
\caption{Calculated electron-phonon coupling constants using two methods---the 
  inverse moment of the many-pole $\alpha^2F(\omega)$ (Eq.~\ref{eq:lambda}) and the
  temperature dependence of the quasiparticle linewidth taken from the spectral 
  functions at large $\varepsilon_k$---and experimental results for comparison.}
\label{tab:lambda}
\begin{ruledtabular}
\begin{tabular}{ l r c c r c c c }
   && \lamMP & \lamLW && \lamEX \\
V  &&  1.174 & 0.899  && \mcds{0.82}{b} & \mcds{1.09}{c} & \mcds{0.80}{d}    \\
Nb &&  1.079 & 0.897  && \mcds{1.04}{b} & \mcds{1.06}{c} & \mcds{1.16}{d}    \\
Pb &&  0.946 & 0.955  && \mcds{1.55}{b} & \mcds{1.48}{c} & \mcds{1.45}{d}    \\
Ta &&  0.909 & 0.809  && \mcds{0.78}{b} & \mcds{0.87}{c} &                   \\
Cu &&  0.155 & 0.126  && \mcds{0.10}{a} & \mcds{0.13}{c} & \mcds{0.08}{d} 
\end{tabular}
\end{ruledtabular}
\raggedright{
\textsuperscript{a}Ref.~\onlinecite{chaikinJLTP77}
\hspace{2mm}
\textsuperscript{b}Ref.~\onlinecite{wolf85} 
\hspace{2mm}
\textsuperscript{c}Ref.~\onlinecite{allenPRB87}
\hspace{2mm}
\textsuperscript{d}Ref.~\onlinecite{brorsonPRL90}
}
\end{table}


Evidence for electron-phonon effects in the spectral function have been measured in a
number of cases. For instance, the value of the mean coupling constant $\lambda$ is 
obtained from the slope of the quasiparticle linewidth $\Gamma\sim2\pi\lambda k_BT$ 
versus temperature.\cite{grimvall81} Thus, calculations of quasiparticle linewidths 
characterize the phonon-contributions to the quasiparticle broadening. Our 
calculated quasiparticle peak FWHM (Fig.~\ref{fig:fwhm}) are comparable to those 
measured experimentally. \cite{mcdougallPRB95,eigurenPRL02,reinertPRL03} Due to the 
redistribution of spectral weight from the quasiparticle peak to the phonon 
satellites at $\varepsilon_k\sim \varepsilon_F$, we use the quasiparticle widths at 
large $\varepsilon_k$ to approximate $\lambda$. Taking copper for example, we find a 
slope of $\approx$0.0680 meV/K, corresponding to $\lambda$ = 0.126.
The calculated $\lambda$ for the metals using the quasiparticle linewidths in 
addition to Eq.~(\ref{eq:lambda}) are given in Table \ref{tab:lambda}, along with 
several experimental results for comparison. Overall, there is decent agreement with 
experiment. The heavier metals show more discrepancy, which is likely an effect of 
the absence of spin-orbit coupling in our simulations.\cite{verstraetePRB08,heid2010}

As another application, the value of electron-phonon coupling is also directly 
related to the superconducting critical temperature $T_c$,
\cite{mcmillanPR68,dynesSSC72} i.e.,
\begin{equation}
\begin{aligned}
\label{eq:Tc} 
  T_c&=\dfrac{\omega_{\rm ln}}{1.20}\exp\left[-\dfrac{1.04\left(1+\lambda\right)}
    {\lambda-\mu^\ast\left(1+0.62\lambda\right)}\right] \\
  \omega_{\rm ln}&\equiv\exp\left[\dfrac{2}{\lambda}\int_0^\infty d\omega
    \dfrac{{\rm ln}(\omega)}{\omega}\alpha^2F(\omega)\right],
\end{aligned}
\end{equation}
where $\mu^\ast$ is the Coulomb pseudopotential, a fitting parameter typically
$\sim$ 0.1-0.2.\cite{morelPR62} The $T_c$ calculated for the non-superconducting 
copper, with $\lambda$ = 0.155 and $\mu^\ast$ = 0.1, is extremely low ($\sim10^{-9}$)
as expected. The other metals give $T_c$ on the correct order of magnitude 
($\sim1-10$ K), though the calculation is sensitive to the choice of Coulomb 
pseudopotential.


\section{SUMMARY AND CONCLUSIONS}
\label{sec:summary}

We have implemented a retarded cumulant (RC) expansion approach to calculate phonon
contributions to electron spectral function. This approach goes beyond the standard
GW approximation to include effects of phonon excitation satellites in the electron 
spectral function. Our calculations show that the phonon-contribution to the 
quasiparticle peak is linearly dependent on temperature. We verify that Migdal's 
theorem is generally satisfied for phonons to high accuracy. Thus the effects of 
vertex corrections leading to deviations between the GW and RC approaches and 
multiple satellites in the spectral function and are generally negligible except at 
very low $T$ ($T\lesssim50$ K) and very strong electron-phonon couplings ($\lambda 
\gtrsim 1$), and would require roughly meV resolution to discern experimentally. The
approach is implemented as part of the \aitops{} workflow tool developed by our 
group.\cite{ai2ps} This hybrid code takes advantage of the capabilities of both 
\abinit{} and \feff{} to generate a number of phonon properties, which include x-ray
Debye-Waller factors, phonon contributions to the electron self-energy and spectral 
function, electron-phonon couplings, as well as estimates of the BCS superconductor
critical temperatures. With an appropriate self-energy, the method presented here can
also be extended to treat insulators and molecular systems.


\begin{acknowledgments}
We thank C.~Draxl, L.~Reining, P.~B.~Allen, G.~Rignanese, X.~Gonze, M.~Bernardi, and
K. Jorissen for useful discussions, and S.~R.~Williams and J.~Vinson for assistance
in code development.
The ABINIT code is a common project of the Universit{\'e} Catholique de Louvain, 
Corning Incorporated and other contributors (URL http://www.abinit.org).
This work was supported in part by DOE grant DE--FG02--97ER45623.

\end{acknowledgments}


\bibliographystyle{apsrev}

\end{document}